\documentstyle[epsf]{mn}

\newcount\eqnno \eqnno=0 
\def\numeqn{\global\advance\eqnno by 1 \eqno(\the\eqnno)}

\title[Substructure]{Substructure in dark matter halos:  Towards a
  model of the abundance and spatial distribution of subclumps}
\author[R. K. Sheth]
{Ravi K. Sheth\\
Department of Physics and Astronomy, University of Pittsburgh, 
3941 O'Hara Street, Pittsburgh PA 15260\\
\smallskip
Email: rks12@pitt.edu\\
}

\date{Submitted to MNRAS 15 April 2003}

\begin{document}

\maketitle

\begin{abstract}
I develop a model for the abundance and spatial distribution of 
dark matter subclumps.  
The model shows that subclumps of massive parent halos formed at 
earlier times than subclumps of the same mass in lower mass parents; 
equivalently, halos in dense regions at a given time formed earlier 
than halos of the same mass in less dense regions.  This may provide 
the basis for interpreting recent observations which indicate that 
the stellar populations of the most massive elliptical galaxies are 
also the oldest.  
\end{abstract} 
\maketitle

\section{Introduction}
In hierarchical clustering models about ten percent of the mass of a 
dark matter halo is expected to be in the form of subclumps which are 
the remnants of objects which fell in and were tidally stripped as 
the halo was assembled.  Whereas models of the abundance of halos as 
a function of mass are quite well developed (Press \& Schechter 1974; 
Sheth \& Tormen 1999; Jenkins et al. 2001; Sheth, Mo \& Tormen 2001), 
models of the subclump abundances are not.  The main goal of this 
paper is to derive a model for the distribution of subclump masses.  
Since it is likely that such subclumps are associated with sites of 
galaxy formation (White \& Rees 1978; Colin et al. 1999) such a model 
is a useful step towards interpretting galaxy luminosity functions.  
The model is developed in Sections~\ref{pois} and~\ref{cont}, and 
Section~\ref{examples} illustrates the results.  
The model presented here may provide a useful framework within which 
to discuss the luminosity function of galaxies, as well as the 
formation of massive elliptical galaxies.  These possibilities are 
discussed in Section~\ref{conclude}, which also highlights 
shortcomings of the model, and outlines possible improvements.

\section{The Poisson model}\label{pois}
Sheth \& Pitman (1997) described a model in which the number density 
of clumps containing $m$ particles is 
\begin{equation}
 n(m,b) = (1-b)\, {(mb)^{m-1}{\rm e}^{-mb}\over m!}, 
 \label{nm}
\end{equation}
where $b$ which grows from an initial value of zero towards an upper 
limit of unity, plays the role of a time variable.  
They showed that the rate of change of the number density of 
$m$-halos could be thought of as the difference between formation 
and destruction rates:
\begin{equation}
 {{\rm d}n(m,b)\over {\rm d}b} =  {{\rm d}n_{\rm form}(b|m)\over {\rm d}b}   
                                - {{\rm d}n_{\rm dest}(b|m)\over {\rm d}b}.  
 \label{dndb}
\end{equation}
where 
\begin{eqnarray}
 {{\rm d}n_{\rm form}(b|m)\over {\rm d}b} &=& {(m-1)\over b}\, n(m,b)\\
 {{\rm d}n_{\rm dest}(b|m)\over {\rm d}b} &=& 
   \left(m + {1\over 1-b}\right)\,n(m,b).
 \label{addcoal}
\end{eqnarray}
In this model, an object of mass $M$ at some late time $b_0$ 
was previously (i.e., at some epoch $b_1\le b_0$) made up of many 
smaller pieces.  The number of $m$-subclumps of an $M$-parent is 
\begin{equation}
 N(m|M,B) = (1-B)\,{M\choose m}\,\left({mB\over M}\right)^{m-1}
 \left(1 - {mB\over M}\right)^{M-m-1}
 \label{NmM}
\end{equation}
where $B=b_1/b_0 \le 1$.  
Extending their model to compute the formation and destruction rates 
of these $m$-subclumps yields:  
\begin{eqnarray}
 {{\rm d}N_{\rm form}(B|m,M)\over {\rm d}B} &=& 
           {(m-1)\over B}\, N(m|M,B)\\
 {{\rm d}N_{\rm dest}(B|m,M)\over {\rm d}B} &=& 
           {(1-m/M)\over (1-mB/M)}\, \\
           && \ \times \left(m + {1\over 1-B}\right)\, N(m|M,B).
 \label{condcoal}
\end{eqnarray}
Details are in the Appendix.  
In the limit that $m\ll M$, equation~(\ref{condcoal}) reduces to 
the unconstrained rates.  As a consistency check, note that 
\begin{equation}
 \sum_{M=m}^\infty {{\rm d}N_{\rm form}(B|m,M)\over b_0\,{\rm d}B}\,n(m,b_0) 
        = {{\rm d}n_{\rm form}(b_1|m)\over {\rm d}b_1},
\end{equation}
where we have used the fact that $dB = db_1/b_0$.  
This shows that the number of $m$-subclumps which form within 
$M$-parent halos during the interval ${\rm d}b_1$ around $b_1$, 
times the density of $M$-parents, when summed over all $M$ does 
indeed give the total formation rate density of $m$-clumps.  
A similar expression holds for the destruction rates.  

In this model, the total number density of $m$ halos which will ever 
form is 
\begin{eqnarray}
 n(m) &=& \int_0^1 {\rm d}b\,{{\rm d}n_{\rm form}(b|m)\over {\rm d}b}
  \nonumber\\
      &=& {(m-1)\over m!} \Bigl[\Gamma(m-1,0,m)
                                - {\Gamma(m,0,m)\over m}\Bigr],
 \label{nmax}
\end{eqnarray}
where the expressions in square brackets denote incomplete gamma 
functions, with parameters $m-1$ and $m$ respectively.  In the limit 
of large $m$, this total number reduces to $1/\sqrt{2\pi m^3}$.

Similarly, integrating $dN_{\rm form}/dB$ over $B$ gives the total 
number of $m$-subclumps which ever formed within an $M$-halo.  
With suitable choices of $m$ and $M$, the value of this integral 
can be used to estimate the typical number of galaxy-sized subclumps 
which are today within virialized clusters.  
Notice that this total number is independent of $b_0$; i.e., it 
depends only on the mass $M$, and not on the time at which the 
parent $M$-halo formed.  

\section{The continuum limit}\label{cont}
Setting $b=1/(1+\delta_{\rm sc})$, and taking $m\gg 1$ and 
$\delta_{\rm sc}\ll 1$ makes $n(m,b)$ resemble the Press-Schechter 
(1974) model of the halo mass function.  Halo abundances in this 
model are usually written as 
\begin{equation}
 {m n(m|t)\over \bar\rho}\,{\rm d}m = {{\rm d}\sigma^2(m)\over \sigma^2(m)}\,
   {\delta_{\rm sc}(t)/\sigma(m)\over \sqrt{2\pi}} 
   \exp\left(-{\delta_{\rm sc}^2(t)\over 2\sigma^2(m)}\right)
 \label{nmps}
\end{equation}
where $\bar\rho$ is the comoving mass density, 
$\delta_{\rm sc}(t)$ is the linear theory overdensity required 
for spherical collapse at $t$ (it is $1.686\,(t_0/t)^{2/3}$ in an 
Einstein-de Sitter model), 
\begin{equation}
 \sigma^2(m) = 4\pi \int {{\rm d}k\over k} k^3P(k) |W(kR)|^2
\end{equation}
is the linear theory variance of the density field at the present 
time $t_0$ when smoothed with a tophat filter on the scale 
$R = (3m/4\pi\bar\rho)^{1/3}$, 
and $P(k)\propto k^n$, with $n=0$ for the Poisson model.  
The use of the tophat filter makes $W(x) = (3/x^3)[\sin(x)-x\cos(x)]$.

The usual Press-Schechter argument is to assume that expressions 
derived from Poisson or white-noise initial conditions are also valid 
for arbitrary Gaussian fluctuation fields (i.e., $P(k)$ may depend 
on $k$).  Therefore, where possible in what follows, we will write 
all expressions in variables which are independent of the power 
spectrum.  

The continuum limit of equation~(\ref{NmM}) shows that 
$mN(m,t|M,T)\,{\rm d}m/M$ is given by an expression like the one 
above, but with the replacements 
$\sigma^2\to \sigma^2(m)-\sigma^2(M)$ and 
$\delta_{\rm sc}\to \delta_{\rm sc}(t)-\delta_{\rm sc}(T)$.  
This is the same as the conditional mass function of 
Bower (1990), Bond et al. (1991) and Lacey \& Cole (1993).  
The continuum limits of the formation time distributions given 
above are 
\begin{equation}
 {{\rm d}n_{\rm form}(t|m)\over {\rm d}t}\,{\rm d}t \to m n(m|t)
   \Big\vert{{\rm d}\delta\over {\rm d}t}\Big\vert\,{\rm d}t
 \label{form}
\end{equation}
and 
\begin{equation}
 {{\rm d}N_{\rm form}(t|m,M,T)\over {\rm d}t}\,{\rm d}t \to 
   m\,N(m,t|M,T)\,
   \Big\vert{{\rm d}\delta\over {\rm d}t}\Big\vert\,{\rm d}t  ,
 \label{formM}
\end{equation}
where $m\le M$ and $t\le T$.  
Furthermore, one can verify that 
\begin{equation}
 \int_m^\infty {{\rm d}N_{\rm form}(t|m,M,T)\over {\rm d}t}
                  \,n(m,T)\,{\rm d}m
  = {{\rm d}n_{\rm form}(t|m)\over {\rm d}t}.
 \label{consistency}
\end{equation}
Expressions similar to equations~(\ref{form}) and~(\ref{formM}) above 
for the formation time distribution have been derived by 
Percival \& Miller (1999) from a quite different approach.  
However, they were only able to describe the shape of the formation 
time distribution, not the normalization.  What we have shown here is 
that the normalization can be derived from the earlier results of 
Sheth \& Pitman (1997).  In hindsight, notice that requiring the 
identity above (equation~\ref{consistency}) to be satisfied also 
provides a way to determine the normalization factors.  

Note that these expressions for the formation times show that 
subclumps of the same mass form at earlier times in massive halos 
than in less massive ones.  We will return to this shortly.  

\subsection{Subclump abundances}
The total number of $m$ subclumps which form within an $M$-halo is 
\begin{equation}
 N_{\rm sub}(m|M,T) = 
    \int_0^T {\rm d}t {{\rm d}N_{\rm form}(t|m,M,T)\over {\rm d}t} .
 \label{nsubmM}
\end{equation}
If these subclumps survive the processes of tidal stripping and 
dynamical friction as they merge into their parent halo (an 
unrealistic approximation!), then the total number density of 
$m$-subclumps at time $T$ is 
\begin{equation}
 n_{\rm sub}(m,T) = \int_m^\infty {\rm d}M\,n(M,T)\,N_{\rm sub}(m|M,T).
 \label{nsubm}
\end{equation}

\subsection{Large-scale spatial distribution of subclumps}
The distribution of $M$-halos is biased with respect to the overall 
distribution of dark matter.  On scales which are large compared to 
the typical diameter of a halo, this bias is factor is approximately 
independent of scale, and is usually denoted $b(M,T)$.  If there are 
$N_{\rm sub}(m|M,T)$ subclumps for each $M$-halo, then, on large 
scales, the bias factor associated with the subclump distribution is 
given by a simple counting argument:  
\begin{equation}
 b_{\rm sub}(m,T) = 
       {\int_m^\infty {\rm d}M\,n(M)\,N_{\rm sub}(m|M,T)\,b(M)\over 
        \int_m^\infty {\rm d}M\,n(M,T)\,N_{\rm sub}(m|M,T)} .
 \label{halobias}
\end{equation} 
The subclump distribution on smaller scales can be estimated if one 
assumes a model for the density run of subclumps around their parent 
halos.  For instance, one might assume that they trace the density 
profile of the mass, or that this distribution is a non-singular isothermal 
sphere.  Once this density profile has been specified, the small scale 
distribution of the subclumps can be described rather compactly 
using the halo model of large scale structure (see the recent review 
in Cooray \& Sheth 2002).  

\subsection{Dependence on local density}
It is sometimes of interest to estimate the distribution of subclumps 
as a function of local density.  We develop two approximations for 
this quantity.  The first is a simple counting argument which exploits 
the fact that the number of $m$-subclumps of an $M$-parent does not 
depend on any quantities other than $m$ and $M$.  If the local 
overdensity $\delta$ is defined on sufficiently large scales that 
$M_\delta\equiv \bar\rho V(1+\delta)$ is much larger than the 
mass of any parent halo, then the density of subclumps in such 
regions is 
\begin{equation}
 n_{\rm sub}(m,T|\delta) = 
   \int_m^{M_\delta} {\rm d}M\,n(M,T|\delta)\,N_{\rm sub}(m|M,T)
 \label{nsubdelta}
\end{equation}
where the density dependent parent halo mass function $n(M,T|\delta)$ 
can be estimated following Mo \& White (1996), 
Kauffmann \& Lemson (1999) and Sheth \& Tormen (2002).  
On very large scales, $M_\delta\gg M$, 
the parent halo mass function reduces to 
$n(M,T|\delta) \to [1 + b(M)\delta]\,n(M,T)$.  
Thus, on these scales, 
 $n_{\rm sub}(m,T|\delta)\to [1 + b_{\rm sub}(m,T)\delta]\,n_{\rm sub}(M,T)$,
with $b_{\rm sub}(m,T)$ given by the expression derived in the 
previous subsection.  On intermediate scales, this simple expression 
for $n(M,T|\delta)$ is no longer valid, but equation~(\ref{nsubdelta}) 
will remain accurate provided that the full expression for $n(M,T|\delta)$
is used.  

Equation~(\ref{nsubdelta}) becomes inaccurate once the cell size $V$ 
on which $\delta$ is evaluated is comparable to the diameter of a 
typical halo.  On such small scales, some cells may only contain 
portions of parent halos, and hence the simple counting argument 
is no longer accurate.  Then it is more accurate to simply compute 
$N_{\rm sub}(m|M_\delta,T_\delta)$, where $T_\delta$ is obtained by 
using the spherical collapse relation for the linear and nonlinear 
overdensities, $\delta_0(\delta)$ (Mo \& White 1996 give a convenient 
fitting formula for this relation), and then set 
$\delta_{\rm sc}(T_\delta) = \delta_0(\delta)$.  Note that, 
in contrast to equation~(\ref{nsubmM}), here the upper limit 
of the integral, $T$, is not the same as the time when the 
parent halo is identified, $T_\delta$.  This is because not all 
the subclumps which will form within $M_\delta$ by $T_\delta$ have 
formed at $T\le T_\delta$.  Thus, 
\begin{equation}
 N_{\rm sub}(m|M_\delta,T_\delta) = \int_0^T {\rm d}t \,
      {{\rm d}N_{\rm form}(t|m,M_\delta,T_\delta)\over {\rm d}t}.
 \label{nsubbias}
\end{equation}
Dividing this quantity by $n_{\rm sub}(m,T)V$ provides an estimate 
of how biased the subclump distribution in dense cells is relative to 
the average.  
Writing the density-dependent abundance this way shows explicitly 
that subhalos of a given mass in dense regions formed at earlier times 
than did subhalos of the same mass in less dense regions.

\section{Explicit results}\label{examples}

\begin{figure}
\centering 
\epsfxsize=\hsize\epsffile{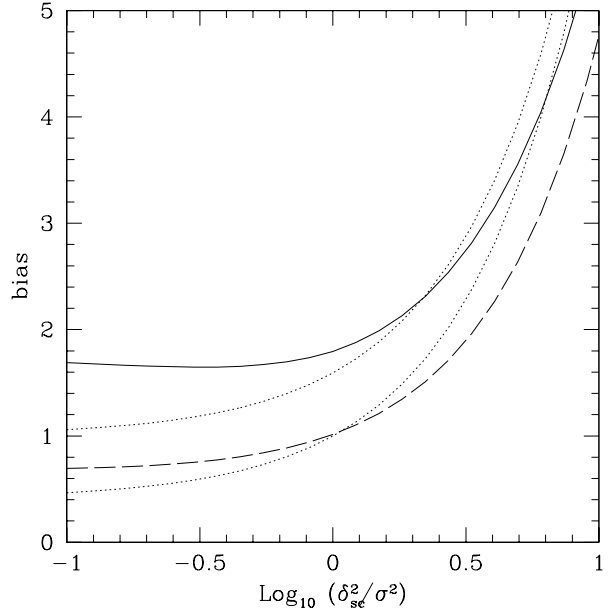}
 \caption{Large-scale bias factor of the parent halos (dashed line, 
 from Sheth \& Tormen 1999) and for subclumps (solid line).  
 The lower and upper dotted curves show the parent and subclump bias 
 factors associated with equations~(\ref{biasmw}) and~(\ref{bsubps}).}
 \label{biassub}
\end{figure}

For the parent halo mass function in equation~(\ref{nmps}), 
all the integrals above can be performed analytically:
\begin{equation}
 N_{\rm sub}(m|M,T)\,{\rm d}m = {M\over \sqrt{2\pi}}
        \,{{\rm d}\sigma^2(m)\over\sqrt{\sigma^2(m)-\sigma^2(M)}},
 \label{NmMT}
\end{equation}
(note that this is independent of $T$) and so the abundance of 
subclumps is 
\begin{equation}
 {n_{\rm sub}(m,T)\over\bar\rho}\,{\rm d}m = 
                 {{\rm d}\sigma^2(m)\over \sqrt{2\pi\sigma^2(m)}} \,
                 \exp\left(-{\delta_{\rm sc}^2(T)\over 2\sigma^2(m)}\right).
 \label{nsubmps}
\end{equation}
The halo bias factor associated with equation~(\ref{nmps}) is 
\begin{equation}
 b(M,T) = 1 + {\delta_{\rm sc}^2(T)/\sigma^2(M) - 1\over\delta_{\rm sc}(T)}
 \label{biasmw}
\end{equation}
(Cole \& Kaiser 1989; Mo \& White 1996; Sheth \& Tormen 1999),
so the large scale bias factor of the subclumps, computed using the 
simple counting argument of equation~(\ref{halobias}) is 
\begin{equation}
 b_{\rm sub}(m,T) = 1 + {\delta_{\rm sc}^2(T)/\sigma^2(m)\over
                          \delta_{\rm sc}(T)}.
 \label{bsubps}
\end{equation}
Whereas the halo bias factor can be less than unity, the bias factor 
of the subclumps is always greater than unity.  The more accurate 
model for subclump biasing (equation~\ref{nsubbias}), which should 
be accurate on smaller scales, can also be solved analytically: 
\begin{eqnarray}
 {N_{\rm sub}(m|M_\delta,T_\delta)\over n_{\rm sub}(m,T)V }
  &=& {(1+\delta)\over\sqrt{1 - \sigma^2(M_\delta)/\sigma^2(m)}}\nonumber\\ 
  & &  \times\quad 
   {\exp[-\nu_\delta^2/2)\over\exp[-\delta_{\rm sc}(T)^2/2\sigma^2(m)]},
\end{eqnarray}
\begin{displaymath}
 {\rm where}\ \ \nu_\delta^2 =
    {\Big[\delta_{\rm sc}(T)-\delta_0(\delta)\Big]^2\over 
                 \sigma^2(m)-\sigma^2(M_\delta)}.
\end{displaymath}
If this ratio is expressed as a Taylor series in $\delta$, then 
the coefficient of the term which is proportional to $\delta$ is 
the same as the linear bias factor derived from the counting argument 
(equation~\ref{bsubps}).  To see this explicitly, consider 
the limit of large cells, for which $M_\delta$ is large simply 
because $V$ is large.  Then $\sigma(M_\delta)/\sigma(m)\to 0$ and 
$\delta_0(\delta)\sim \delta \ll \delta_{\rm sc}$, so that 
\begin{equation}
 {N_{\rm sub}(m|M_\delta,T_\delta)\over n_{\rm sub}(m,T)V}
  \to (1+\delta)\,\left[1 + 
  \delta\,{\delta^2_{\rm sc}(T)/\sigma^2(m)\over \delta_{\rm sc}(T)}\right].
\end{equation}

Numerical simulations show that the spherical collapse based 
equation~(\ref{nmps}) is a good but not perfect description of the 
number density of parent halos.  A more accurate formula is 
\begin{equation}
 {m n(m|t)\over \bar\rho}\,{\rm d}m = 
   {{\rm d}\nu^2\over \nu^2}\,
   \sqrt{a\nu\over 2\pi} \, \exp\left(-{a\nu^2\over 2}\right)
   A\,\left[1 + (a\nu)^{-2p}\right] 
 \label{st99}
\end{equation}
(Sheth \& Tormen 1999), where 
$\nu = \delta_{\rm sc}(t)/\sigma(m)$, 
$a\approx 0.71$, $p=0.3$ and $A=0.322$ insures that the distribution 
is normalized to unity.  Sheth, Mo \& Tormen (2001) argue that 
this expression may be related to models in which halos form from an 
ellipsoidal collapse.  
Percival, Miller \& Peacock (2000) show that the insertion of 
equation~(\ref{st99}) in equation~(\ref{form}) provides a better 
description of halo formation in simulations than does 
equation~(\ref{nmps}).  Therefore, it probably provides a more 
accurate model of the subclump abundances and bias factors as well.  
To use this requires a model for the analog of the conditional 
mass function (equation~\ref{NmM}); Sheth \& Tormen (2002) argue 
that simply changing variables in $n(m,t)$, as is appropriate when 
equation~(\ref{nmps}) is the mass function, while incorrect, is 
not a bad approximation.  In this approximation, 
the analog of equation~(\ref{NmMT}) 
is again proportional to $M/\sqrt{\sigma^2(m)-\sigma^2(M)}$, 
the subclump mass function (integrated over parent masses $M$) 
can be written in terms of hypergeometric functions, as can 
the large-scale bias factor.  The expressions are lengthy, 
so I haven't reproduced them here.  Figure~\ref{biassub} compares 
the large-scale bias factor of the parent halos (dashed, from 
Sheth \& Tormen 1999), with that for the subclumps (solid).  
The lower and upper dotted curves show the parent and subclump bias 
factors associated with equations~(\ref{biasmw}) and~(\ref{bsubps}).  

\section{Discussion}\label{conclude}
I derived a model of the subclump distribution under the assumption 
that there are no processes by which a subclump can lose mass as it 
falls in to a larger system.  Although this idealization is not 
realistic, it should be thought of as providing the initial conditions 
for more sophisticated calculations which do incorporate the effects 
of tidal stripping and dynamical friction.  
For instance, the cumulative distribution of subclumps in this 
model is 
\begin{displaymath}
  \int_m^M N_{\rm sub}(m|M,T)\,{\rm d}m = {2M\sigma(m)\over \sqrt{2\pi}}
                                    \,\sqrt{1-{\sigma^2(M)\over\sigma^2(m)}}.
\end{displaymath}
When $m\ll M$, this scales as $M\,\sigma(m)$, which is rather 
different from the $m^{-1}$ scaling seen in simulations after 
tidal stripping.  
Nevertheless, the model is useful because it provides simple 
closed-form expressions for how the abundances and spatial distributions 
of the subclumps differ from those of their parent halos.  

If mergers are unimportant, then equation~(\ref{nsubmps}) can be 
turned into an estimate of the luminosity function.  This can be 
done either by assuming a mean mass-to-light ratio, or by assuming 
some relation for how the light-curve evolves with time, and then 
convolving with the formation time distributions derived here.  
I have not pursued this further because this model for the subclump 
distribution is not sufficiently realistic.  For example, 
$\int {\rm d}m\,mn_{\rm sub}(m)/\bar\rho$ is generally greater than 
unity.  This is a consequence of the fact that objects counted as 
having formed with mass $m$ may previously have been counted as 
having formed with mass $m'<m$.  A more realistic model of the 
subclump distribution would avoid this double counting:  one 
attractive model in this regard is to study how the mass of the 
most massive progenitor subclump decreases with lookback time.  
One then labels as subclumps all the objects which merge with the 
most massive progenitor.  This distribution of subclumps can be 
described following results in Nusser \& Sheth (1999) and is the 
subject of work in progress.  

\begin{figure}
\centering 
\epsfxsize=\hsize\epsffile{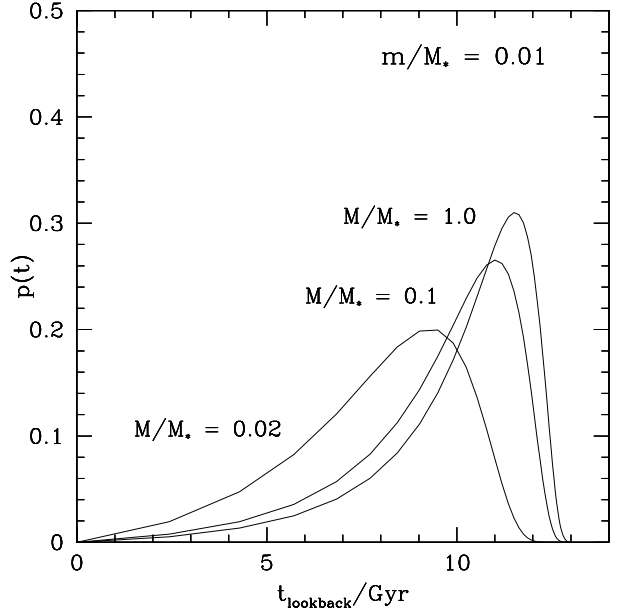}
 \caption{Distribution of formation times for subclumps of 
 mass $m/M_*=0.01$ for various choices of the parent mass $M$ 
 at $z=0$.  A power-law spectrum with $n=-2$ was used; 
 lookback times were computed assuming a flat cosmology with 
 $\Omega_0=0.3$, $\Lambda_0=1-\Omega_0$ and Hubble constant 
 $H_0=70$~km~s$^{-1}$~Mpc$^{-1}$.  To set the scale, the value 
 of $M_*$ in such a $\Lambda$CDM model would be 
 $\sim 10^{13} h^{-1}M_\odot$.  }
 \label{ptdt}
\end{figure}

There is one respect in which the model developed here is realistic:  
it shows that subclumps of massive parent halos formed at earlier 
times than subclumps of the same mass in lower mass parents, 
or equivalently, that halos in dense regions at a given time 
formed earlier than halos of the same mass in less dense regions.  
This has an interesting consequence for the following simple model 
of elliptical galaxies.  Suppose that gas can only cool and form 
stars within halos which are more massive than some minimum mass $m$.  
If this happened at high redshift, then the halos within which the 
stars formed will have subsequently merged with other halos to make 
more massive objects.  If we treat the separate parcels of stars as 
representing the subclumps $m$, then we have a model in which massive 
galaxies contain older stars.  Moreover, since higher redshifts 
correspond to smaller intervals of time, the distribution of formation 
times will be narrower for the stars which form in more massive halos, 
and broader for the stars which form in lower mass halos; 
Figure~\ref{ptdt} illustrates.  
This is in qualitative agreement with recent work 
(Thomas, Maraston \& Bender 2002).

\section*{Acknowledgments}
Most of the results presented here were obtained while I was a postdoc 
at Fermilab, supported by the DOE and NASA grant NAG 5-10842.  
I'd like to thank Scott Dodelson, Josh Frieman, Albert Stebbins 
and Liz Duty for making Fermilab a rewarding place to work.

\appendix
\section{The additive coagulation model}
In the model discussed in the main text, halos grow by binary 
mergers with one another.  The formation rate of a halo with 
$M$ particles is given by evaluating 
\begin{displaymath}
 {M\over 2} \sum_{m=1}^{M-1} n(m,b) n(M-m,b),
\end{displaymath}
which expresses the fact that formation in this model happens by 
binary mergers, with probability of merger proportional to the mass 
of the merging pair, and the abundance of each type of clump.  
The associated destruction rate is related to 
\begin{displaymath}
 n(M,b) \sum_{m=1}^\infty (M+m)\, n(m,b).  
\end{displaymath}
This system of equations for $n(M,b)$ is solved by equation~(\ref{nm}); 
this was one of the results in Sheth \& Pitman (1997).

The conditional formation rate of $m$-subclumps within an $M$-parent, 
while given by a similar expression, is slightly different.  
It is  
\begin{displaymath}
 {m\over 2} \sum_{j=1}^{M-m-1} N(j,b_1|M,b_0) N(m-j,b_1|M-j,b'),
\end{displaymath}
where $1/b' = (M-j)/(Mb_0-jb_1)$.  
The first term is the number of $j$-subclumps within the $M$-halo, 
and the second term is the number of subclumps containing $m-j$ 
particles, given that there are only $M-j$ remaining particles from 
which they could have been formed (see Sheth \& Lemson 1999 for a 
detailed discussion); this is why the second term is not 
$N(m-j,b_1|M,b_0)$.  
For similar reasons, the destruction rate is obtained from 
\begin{displaymath}
 N(m,b_1|M,b_0) \sum_{j=1}^{M-m-1} (m+j)\, N(j,b_1|M-m,b')
\end{displaymath}
rather than 
$N(m,b_1|M,b_0) \sum_{j=1}^{M-m-1} (m+j) N(j,b_1|M,b_0)$.  

\end{document}